\documentclass[11pt,a4paper]{article}

\usepackage{amsmath}
\usepackage{amsfonts}
\usepackage{mathrsfs}
\usepackage[T1]{fontenc}
\usepackage{graphicx}
\usepackage{color}

\newcommand{\beq}{\begin{equation}}
\newcommand{\eeq}{\end{equation}}
\newcommand{\bal}{\begin{aligned}}
\newcommand{\eal}{\end{aligned}}
\newcommand{\scri}{\mathscr{I}}
\newcommand{\D}{\mathrm{d}}

\begin{document}
\frenchspacing

\title{A Toy Penrose Inequality and Its Proof}
\author{Ingemar Bengtsson \\
 Emma Jakobsson}
\date{\it \small Stockholms Universitet, AlbaNova \\
Fysikum \\
S-106 91 Stockholm, Sweden}
\maketitle

\begin{abstract}
\noindent We formulate and prove a toy version of the Penrose inequality.
The formulation mimics the original Penrose inequality in which the scenario is the following: A shell of null dust collapses in Minkowski space and a marginally trapped surface forms on it.
Through a series of arguments relying on established assumptions, an inequality relating the area of this surface to the total energy of the shell is formulated.
Then a further reformulation turns the inequality into a statement relating the area and the outer null expansion of a class of surfaces in Minkowski space itself.

The inequality has been proven to hold true in many special cases, but there is no proof in general.
In the toy version here presented, an analogous inequality in (2+1)-dimensional anti-de Sitter space turns out to hold true.
\end{abstract}


\section{Introduction}
This story began in 1973, when Penrose formulated an inequality which, if violated, would suggest that at least one of a collection of established assumptions about gravitational collapse---one of which is cosmic censorship---must be false \cite{penrose73}.
Imagine that an infinitesimally thin shell of null dust collapses in flat spacetime, and that a marginally trapped surface forms on it.
Then the inequality states that
\beq \label{eq:pi}
	A \leq 16\pi M^2,
\eeq
where $A$ is the area of the trapped surface, and $M$ is the total energy of the null shell.
The arguments leading up to the inequality require that the shell has no caustics to the past of the surface.
This ensures that the entire interior of the shell is truly Minkowski space, that there are no curvature singularities in the past, and that an asymptotically defined mass can be assigned to the shell.
The only restriction on the exterior spacetime is that no further matter or radiation is falling in after the shell has passed.
Apart from that its detailed properties are not relevant.
It should be mentioned though that the inequality is saturated if the resulting spacetime geometry is Schwarzschild and the marginally trapped surface is the round sphere of the event horizon.
In that case the total energy $M$ must be equal to the Schwarzschild mass.

By adjusting the mass density of the shell properly, any spacelike cross-section can be made into a marginally trapped surface.
This maneuver turns the inequality into a statement about the geometry of Minkowski space itself.
In his attempt to violate the inequality, Penrose let the null shell have the geometry of a backwards light cone of a point.
Then he studied an arbitrary spacelike cross-section of the shell and ended up with an inequality which he could not find a counterexample to.
Later it was shown that this particular inequality is indeed true \cite{tod85}.
Another special case of the inequality was early on shown to hold true by Gibbons \cite{gibbons73}.
He considered null shells of more general shape, but let the marginally trapped cross-section lie in a flat spacelike hyperplane.
In this case, the condition that the null shell is free of caustics in the past turns into a condition on the cross section to be convex.
The Penrose inequality then follows from the Minkowski inequality \cite{minkowski03}, which in turn is related to the isoperimetric inequality relating the area of a closed surface to the volume contained in it.\footnote{See for instance \cite{tod92} for an overview.}

Since then inequalities of the form \eqref{eq:pi} have been studied in different contexts, with careful specifications of what surface $A$ is the area of, and which mass $M$ is \cite{mars09}.
One formulation of the inequality, where decisive progress has been made, is when considering an asymptotically flat initial data set.
Letting $A$ be the area of an appropriately defined minimal surface on the set, and $M$ be the ADM mass, the inequality \eqref{eq:pi} has been proven when the initial data set is time-symmetric---the {\it Riemannian Penrose inequality}.
Some early partial results were obtained by Jang and Wald \cite{jang77}, building on work by Geroch \cite{geroch73}.
Much later this version of the inequality was proven in full generality \cite{huisken01,bray01}.

Lately progress has been made in the original setup \cite{mars12,mars14}.
The original idea has also been generalized to other spacetime geometries---including Minkowski as a special case---for instance Schwarzschild spacetime \cite{brendle14}, and asymptotically flat spacetimes satisfying the dominant energy condition in general \cite{mars15}.
Taken together all of these results have shown that the inequality holds for a large class of surfaces.
But a full proof has not yet been put forward.\footnote{A full proof was claimed in \cite{gibbons97}, but it contains an error, see \cite{mars09,mars12}.}

In this paper we write down a Penrose-like inequality in (2+1)-dimensional anti-de Sitter space and prove it.
The outline of the paper is the following:
In Section \ref{sec:statement} the original inequality and how it reduces to a geometric property of surfaces in Minkowski space is reviewed.
The toy version in anti-de Sitter space is described in Section \ref{sec:toy} and proven in Section \ref{sec:proof}.
In Section \ref{sec:summary} we give a summary and concluding remarks.


\section{A Statement About Minkowski Space} \label{sec:statement}
Let us first restate the Penrose inequality in the following form:
\beq \label{eq:pi2}
	M \geq \sqrt{\frac{A}{16\pi}}.
\eeq
The left hand side, it turns out, can be expressed in terms of the outer null expansion the marginally trapped surface would have had if it had been a surface in pure Minkowski space.

How the null expansion is related to the energy of the shell can be understood in the following way:
Consider an expanding null hypersurface in the flat region intersecting the collapsing null shell along the surface in question.
As this expanding family of null geodesics passes the shell it will tend to be focussed by an amount determined by the energy density of the collapsing null dust.
So the outer null expansion will make a jump across the shell, and the energy density of the shell can be chosen so that the jump becomes exactly equal to the null expansion of the surface on the Minkowski side, thus making the surface marginally outer trapped.

A sketch of the technical details is given in the following.
Consider a null hypersurface $\Sigma$ with future-directed normal vector field $\vec{k}$, dividing spacetime into an exterior region $V^\text{ext}$ and an interior region $V^\text{int}$ with corresponding metrics $g^\text{ext}_{ab}$ and $g^\text{int}_{ab}$.
We can write down the full spacetime metric as
\beq
	g_{ab} = \Theta \,g^\text{ext}_{ab} + (1-\Theta) g^\text{int}_{ab},
\eeq
where $\Theta$ is the Heaviside step function:
\beq
	\Theta = \begin{cases}
		1 & \quad \text{in } V^\text{ext} \\
		1/2 & \quad \text{on } \Sigma \\
		0 & \quad \text{in } V^\text{int}
	\end{cases}
\eeq
We assume that the metric is continuous across $\Sigma$, i.e. $\left.g^\text{ext}_{ab}\right|_\Sigma = \left.g^\text{int}_{ab}\right|_\Sigma$.
Then, writing down a tensor distribution associated to $g_{ab}$ we find that the corresponding stress-energy tensor distribution on $\Sigma$, in the case of null dust, is
\beq
	\boldsymbol{T}_{ab} = \boldsymbol{\delta} \mu \, k_a k_b.
\eeq
The energy density $\mu$ can be expressed in terms of the jumps of the derivatives of the metric.\footnote{It also depends on the normalization of $\vec{k}$. Under a rescaling $\vec{k} \to \sigma \vec{k}$ the energy density scales as $\mu \to \sigma^{-1}\mu$. More details on the stress-energy tensor for null shells can be found in the literature, e.g. \cite{mars93,poisson04}.}
The $\delta$-distribution is defined as
\beq
	\boldsymbol{\delta} k_a = \nabla_a \boldsymbol{\Theta},
\eeq
with $\boldsymbol{\Theta}$ being the scalar distribution corresponding to the Heaviside step function.

When there is a static Killing vector field $\vec{\xi}$, there is a conserved current
\beq
	\boldsymbol{J}^a = -\boldsymbol{T}^{ab} \xi_b
\eeq
along the null shell.
The corresponding conserved charge $M$ can be evaluated as
\beq \label{eq:mass}
	M = \oint_{S} \mu \, (-\vec{k} \cdot \vec{\xi}) \, \mathrm{d}S
\eeq
where $S$ is any spacelike cross section of the shell.
This is precisely the Bondi mass evaluated at the intersection of the null shell with past null infinity.

Now, choose any spacelike cross section $S$ of the null shell.
At every point of this surface there are two future-directed null normal vectors, one of which is $\vec{k}^- = \vec{k}$.
This is the same as the normal to the collapsing shell and it is inward directed.
Denote the outward directed null normal by $\vec{k}^+$ and choose its normalization so that
\beq \label{eq:normalk+}
	\vec{k}^+ \cdot \vec{k}^- = -2.
\eeq
The above condition does not completely fix the null normals since Eq. \eqref{eq:normalk+} will still hold under a rescaling $\vec{k}^- \to \sigma\vec{k}^-$ and $\vec{k}^+ \to \vec{k}^+/\sigma$.
This ambiguity will later be settled by the condition
\beq \label{eq:normalk-}
	\vec{k}^- \cdot \vec{\xi} = -1.
\eeq	
Evaluating the energy density on $S$ it is found that
\beq \label{eq:mu}
	\mu = \frac{1}{16\pi}(\theta_+^\text{int} - \theta_+^\text{ext}),
\eeq
where $\theta_+$ is the outer null expansion of the surface, which will differ depending on which side of the null the shell it is evaluated from.
The surface will be marginally outer trapped if we choose $\mu$ so that $\theta_+^\text{ext}=0$ everywhere along it.
With this choice of the energy density, inserting Eq. \eqref{eq:mu} in the expression \eqref{eq:mass} of the mass we find that the inequality \eqref{eq:pi2} takes the form
\beq \label{eq:pitheta}
	\oint_S \theta_+^\text{int} \, (-\vec{k}^- \cdot \vec{\xi}) \, \mathrm{d}S \geq \sqrt{16\pi A}.
\eeq
This is the form of the Penrose inequality that we put our emphasis on in this paper.
Note the remarkable fact that the above inequality contains properties of the surface $S$ on the Minkowski side of the shell only.
And we expect it to hold for any closed spacelike surface in flat spacetime as long as the null geodesics in the direction of $-\vec{k}^-$ do not cross anywhere when traced back in time from the surface.

Since we are dealing with a geometric property of surfaces, one may wonder if inequalities similar to \eqref{eq:pitheta} hold true in other contexts.
Given the difficulty to prove it in general, what if we were to lower the dimension?
And what about other spacetime geometries?
These questions give motivation to the toy version of the Penrose inequality presented in the next section.


\section{The Toy Version} \label{sec:toy}
The aim of this section is to formulate a Penrose-like inequality in 2+1 dimensions.
In Minkowski spacetime such a formulation already exists as a special case of a generalization of the inequality \eqref{eq:pitheta} in $n+1$ dimensions \cite{gibbons97}.
It then takes the form
\beq
		\oint_\gamma \theta_+ \, (-\vec{k}^- \cdot \vec{\xi}) \, \mathrm{d}l \geq 2\pi,
\eeq
where the integral is performed over a closed curve $\gamma$, rather than a closed surface.
This inequality does in fact hold true for the relevant class of curves, as can be shown using Theorem 2 in \cite{mars12}.
However, it does not contain the area (or rather the length) of the curve---in accordance with the fact that mass is dimensionless in this context.
Thus we have lost one of the two key ingredients, and we would like to restore it.

Going back to the beginning, we recall that the original idea was to have a collapsing null shell forming a black hole, and that the inequality is saturated in the case of spherical symmetry.
For black holes to exist in 2+1 dimensions we need to add a negative cosmological constant, and thus we turn our attention to anti-de Sitter space.
There the original setup can be mimicked in detail.
We know that a collapsing null fluid can give rise to various black hole solutions \cite{husain94,chan96}; there are some particularly interesting new results where the outcome of an inhomogeneous energy distribution on a light cone is studied in detail \cite{lindgren15}.
We will, however, begin by dealing with the simplest case.
When the shell has the shape of the past light cone of a point and the energy distribution is circularly symmetric it is easy to see that the resulting spacetime is the analogue of the Schwarzschild solution---namely the nonrotating BTZ black hole \cite{banados92}.

On the one hand, the mass $M$ and the length $L_{BTZ}$ of the event horizon in this geometry is related by
\beq
	\ell^2 M = \left( \frac{L_{BTZ}}{2\pi} \right)^2,
\eeq
where $\ell= (-\Lambda)^{-1/2}$ is the natural length scale set by the cosmological constant $\Lambda$.
Inspired by the form \eqref{eq:pi2} of the Penrose inequality this suggests that
\beq \label{eq:MgeqL}
	\ell^2 M \geq \left( \frac{L}{2\pi} \right)^2,
\eeq
for the length $L$ of any marginally trapped curve on a collapsing shell of null dust.
As in the original inequality we require that the null shell is free of caustics to the past of the curve, thus making sure that the shell's interior is anti-de Sitter space.

On the other hand, the event horizon of the exterior BTZ solution intersects the null shell along a circular cross section with circumference $L_{BTZ}$.
Calculating the outer null expansion $\theta^+$ of a circular cross section of the past light cone on the anti-de Sitter side of the shell and integrating over the circle itself we find that
\beq \label{eq:circle}
	\frac{1}{2\pi}\oint \theta^+ \mathrm{d}l = 1 + \left( \frac{L_{BTZ}}{2\pi\ell} \right)^2 = 1 + M .
\eeq
Here the null normal vectors of the circle have been normalized according to Eqs. \eqref{eq:normalk+} and \eqref{eq:normalk-}, using a static Killing vector $\vec{\xi}$.
Combined, Eqs. \eqref{eq:MgeqL} and \eqref{eq:circle} now suggest the inequality
\beq \label{eq:toypi}
	\frac{1}{2\pi}\oint \theta^+ \mathrm{d}l \geq 1 + \left( \frac{L}{2\pi\ell} \right)^2
\eeq
as the analogue of Eq. \eqref{eq:pitheta} in (2+1)-dimensional anti-de Sitter space.

Just as Eq. \eqref{eq:pitheta} is a claim about surfaces in Minkowski space, Eq. \eqref{eq:toypi} is a claim about surfaces in pure anti-de Sitter space.
Therefore no further reference will be made to black hole geometries.
The original inequality is connected to the question of cosmic censorship, but we will not attempt to formulate a cosmic censorship hypothesis for 2+1 dimensions.
With the arena being anti-de Sitter space it will simply be shown in the next section that the toy version \eqref{eq:toypi} of the Penrose inequality does indeed hold true.


\section{The Proof} \label{sec:proof}
Before getting to the proof of the Penrose-like inequality \eqref{eq:toypi} some acquaintance with (2+1)-dimensional anti-de Sitter space is needed.
It can be defined as the hypersurface
\beq
	X^2 + Y^2 - U^2 - V^2 = -\ell^2
\eeq
embedded in a flat spacetime with line element
\beq
	\D s^2 = \D X^2 + \D Y^2 - \D U^2 - \D V^2.
\eeq
From now on $\ell$ will be set to one.
The coordinates we will use are given by
\beq
	\begin{array}{l l}
		X = \frac{2\rho}{1-\rho^2}\cos\phi & \quad Y = \frac{2\rho}{1-\rho^2}\sin\phi \\
		U = \frac{1+\rho^2}{1-\rho^2}\cos t & \quad V = \frac{1+\rho^2}{1-\rho^2}\sin t
	\end{array}
\eeq
in which the line element takes the form
\beq \label{eq:adS}
	\D s^2 = -\left(\frac{1+\rho^2}{1-\rho^2}\right)^2 \D t^2 + \frac{4}{(1-\rho^2)^2}\left(\D \rho^2 + \rho^2 \D \phi^2\right).
\eeq
There is a static Killing vector field
\beq \label{eq:killing}
	\xi^a = \partial_t^a,
\eeq
and slices of constant $t$ are Poincar\'e disks on which $\rho$ is a radial coordinate and $\phi$ is an angular coordinate.
There is a conformal boundary $\scri$ at $\rho=1$.
The advantage of these coordinates---as we will see---is that we can do calculations directly on $\scri$ by multiplying the metric \eqref{eq:adS} by a conformal factor.

In the standard approach to tackling the original Penrose inequality the starting-point is often a surface.
Then comes the question of whether the past expanding null hypersurface generated by null normals to this surface is free of caustics or not.
As mentioned in the introduction, Gibbons found that a surface in a flat spacelike hyperplane fulfils the requirement if it is convex, but for a general surface the question is more difficult to answer.
Similarly, in our context, a curve on the Poincar\'e disk has the desired properties if it is geodetically convex (in the strong sense that the curvature is everywhere positive).
We will, however, follow a slightly different route in our proof.
The strategy is given as follows.
Let an arbitrary collapsing null shell be represented by a null surface defined in terms of its past intersection with $\scri$.
Eventually caustics will appear, but by letting the null surface be defined by a smooth spacelike (achronal) cut of $\scri$ we make sure that for any cross section taken before these appear the null surface will be free of caustics to the past of that cross section.
A general curve of interest is thus defined by two functions: one function on $\scri$ describing the null surface, and one function describing a cross section of the null surface.
By explicitly expressing the outer null expansion and the length of the curve in terms of these two functions we will see that the inequality \eqref{eq:toypi} holds true.

Let us begin with the description of the null surface.
Multiply the metric \eqref{eq:adS} by a conformal factor
\beq \label{eq:omega}
	\Omega = \frac{1-\rho^2}{1+\rho^2},
\eeq
in order to obtain the optical metric \cite{abramowicz88}
\beq \label{eq:optical}
	\D \hat{s}^2 = \Omega^2 \D s^2 = -\D t^2 + \frac{4}{(1+\rho^2)^2}\left( \D\rho^2 + \rho^2\D\phi^2\right).
\eeq
Choose a curve $\gamma_\scri$ on $\scri$ parametrized by $\sigma$ as
\beq
	\gamma_\scri: \quad
	\begin{cases}
		t = T(\sigma) \\
		\rho = 1 \\
		\phi = \sigma 
	\end{cases}
	\quad 0 \leq \sigma \leq 2\pi
\eeq
We will require that the curve is closed and smooth, i.e. that the function $T(\sigma)$ is smooth and periodic with period $2\pi$.
The curve should also be spacelike---a requirement fulfilled if
\beq
	T'^2(\sigma) < 1, \quad \forall \sigma \in [0,2\pi],
\eeq
where the prime denotes differentiation with respect to $\sigma$.
The null surface $\Sigma$ will now be swept out by null geodesics emanating orthogonally from $\gamma_\scri$ and inward directed.
To find a description of these geodesics we note the following: Since the optical metric \eqref{eq:optical} is related to the metric \eqref{eq:adS} of anti-de Sitter space through a conformal factor, null geodesics with respect to the optical metric are also null geodesics with respect to the anti-de Sitter metric.
Furthermore, given the form of the optical metric, null geodesics projected down to space curves at constant $t$ are geodesics with respect to the spatial metric.
And the spatial metric of \eqref{eq:optical} is the metric of the sphere as given by a stereographic projection of its embedding in Euclidean space.
The problem of finding null geodesics in anti-de Sitter space thus reduces to that of finding geodesics on an ordinary sphere.
Working out the details we find that $\Sigma$ is given by
\beq \label{eq:shell}
	\Sigma: \quad
	\begin{cases}
		t(\sigma,\tau) = T(\sigma) + \tau \\
		\rho(\sigma,\tau) = \sqrt{\frac{1-\sqrt{1-T'^2(\sigma)}\sin\tau}{1+\sqrt{1-T'^2(\sigma)}\sin\tau}} \\
		\phi(\sigma,\tau) = \arctan\left(\frac{\sin\sigma\cos\tau + T'(\sigma)\cos\sigma\sin\tau}{\cos\sigma\cos\tau - T'(\sigma)\sin\sigma\sin\tau}\right)
	\end{cases}
	\quad
	\begin{aligned}
		& 0 \leq \sigma \leq 2\pi \\
		& 0 \leq \tau \leq \pi
	\end{aligned}
\eeq
Putting $\tau=0$ in the above expressions we recover the curve $\gamma_\scri$.
Curves of constant $\sigma$ are the null geodesics parametrized by $\tau$.
With the above description of the null surface $\Sigma$ we can leave the optical geometry behind, and from here on work directly in anti-de Sitter space.

Now, we ask ourselves where the caustics on $\Sigma$ first appear.
Consider curves of constant $\tau$ foliating the null surface.
Such a curve---parametrized by $\sigma$---fails to be regular where its tangent vector vanishes, that is when
\beq
	\partial_\sigma x^a(\sigma,\tau) = 0,
\eeq
with $(x^0,x^1,x^2) = (t,\rho,\phi)$.
This happens when $\tau=\tau_c$ and $\sigma=\sigma_c$ such that
\beq \label{eq:caustic}
	\begin{aligned}
		& T'(\sigma_c) = 0, \\
		& \cot(\tau_c) = -T''(\sigma_c).
	\end{aligned}
\eeq
Thus we see that a caustic first appears along a null geodesic where the value of $\sigma$ is such that the function $T(\sigma)$ has a maximum, at a value of $\tau$ determined by the sharpness of this maximum.

Now that we are in control over the null surface and its caustics it is time to analyze a general cross section $\gamma$ of it.
Let $\tau$ be a smooth function of $\sigma$ describing the curve:
\beq
	\gamma: \quad \tau = \tau(\sigma) \quad \text{on } \Sigma.
\eeq
Then the line element on the curve is
\beq
	\D l^2 = A^2(\sigma) \D \sigma^2,
\eeq
where
\beq
	A(\sigma) = \cot[\tau(\sigma)] + \frac{T''(\sigma)}{1-T'^2(\sigma)}.
\eeq
Note that the function $A(\sigma)$ vanishes at the point \eqref{eq:caustic} where a caustic first appears.
Given the line element on the curve we find that its length is
\beq \label{eq:L}
	L = \int_0^{2\pi} \cot[\tau(\sigma)] \, \D\sigma.
\eeq
The tangent vector to the curve is 
\beq
	u^a = t'\partial_t^a + \rho' \partial_\rho^a + \phi' \partial_\phi^a,
\eeq
evaluated on $\gamma$ so that $t(\sigma)$, $\rho(\sigma)$, and $\phi(\sigma)$ are given by Eq. \eqref{eq:shell} with $\tau$ as a function of $\sigma$.
The ingoing null normal vector---which is also tangent to $\Sigma$---is given by
\beq
	k_a^- = -\nabla_a t - \frac{1}{\rho}\sqrt{1-T'^2}\cos\tau\nabla_a\rho + T'\nabla_a\phi.
\eeq
It is normalized in accordance with Eq. \eqref{eq:normalk-} given the Killing vector field \eqref{eq:killing}.
The outgoing null normal to $\gamma$ is given by
\beq
	k_a^+ = \frac{1}{(1-T'^2)\sin^2\tau}\left( -(1+B^2)\nabla_a t + \alpha \nabla_a \rho + \beta \nabla_a \phi \right),
\eeq
where
\beq
	\begin{aligned}
		& B = \frac{t'}{A\sqrt{1-T'^2}\sin\tau} \\
		& \alpha = \frac{1}{\rho}\left( (1-B^2)\sqrt{1-T'^2}\cos\tau + 2BT'\right) \\
		& \beta =  T'(B^2-1) + 2B\sqrt{1-T'^2}\cos\tau
	\end{aligned}
\eeq
The normalization of $\vec{k}^+$ is set so that Eq. \eqref{eq:normalk+} is fulfilled.

Now it remains to express the null expansions of $\gamma$ in terms of the two functions $T(\sigma)$ and $\tau(\sigma)$ defining it.
These are found by evaluating
\beq
	\begin{aligned}
		\theta^\pm & = -\frac{1}{A^2} \, \vec{k}^\pm \cdot \nabla_{\vec{u}}\,\vec{u} \\
		& = -\frac{1}{A^2} \, k^\pm_a u^b \left( \partial_b u^a + {\Gamma^a}_{bc} u^c \right)
	\end{aligned}
\eeq
given the above expressions for the null normal and tangent vectors.
This is a tedious task since we are summing a large number of terms which, individually, do not have any invariant meaning.
After some work one finds
\beq
	\theta^- = -\frac{1}{A}(1-T'^2),
\eeq
and
\beq
	\begin{aligned}
		\theta^+ = & \frac{1}{A} \left[ 1 + \cot^2\tau + \frac{t'^2}{A^2(1-T'^2)\sin^4\tau} \right] \\
		& + \frac{1}{A}\left[ \frac{\D}{\D\sigma} \left( \frac{2t'}{A(1-T'^2)\sin^2\tau} \right) \right].
	\end{aligned}
\eeq
And the work pays off.
With the above expression for the outer null expansion we can evaluate the left hand side of the inequality \eqref{eq:toypi}:
\beq \label{eq:almost}
	\frac{1}{2\pi}\oint_\gamma \theta^+ \, \D l = 1 + \int_0^{2\pi} \cot^2\tau \, \frac{\D\sigma}{2\pi} + \int_0^{2\pi} \frac{t'^2}{A^2(1-T'^2)\sin^4\tau} \, \frac{\D\sigma}{2\pi}.
\eeq
The proof now falls out almost immediately after a little closer inspection of the second and third term on the right hand side of the above expression.
Recalling that the length of $\gamma$ is given by the expression in Eq. \eqref{eq:L} we see that it is related to the second term by
\beq
	\int_0^{2\pi} \cot^2\tau \, \frac{\D\sigma}{2\pi} \geq \left(\int_0^{2\pi} \cot\tau \, \frac{\D\sigma}{2\pi}\right)^2 =  \left(\frac{L}{2\pi}\right)^2.
\eeq
This observation together with the fact that the third term is positive completes the proof, since it then follows that
\beq
	\frac{1}{2\pi}\oint_\gamma \theta^+ \, \D l  \geq 1 + \left(\frac{L}{2\pi}\right)^2.
\eeq
The na\"ive guess \eqref{eq:toypi} obtained by simply copying the logic of Penrose's original inequality thus turns out to be a true statement about (2+1)-dimensional anti-de Sitter space.

\section{Summary and conclusions} \label{sec:summary}
We have shown that
\beq \label{eq:toypi2}
	\frac{1}{2\pi}\oint \theta^+ \mathrm{d}l \geq 1 + \left( \frac{L}{2\pi\ell} \right)^2
\eeq
is a true inequality for a class curves in (2+1)-dimensional anti-de Sitter space with length scale $\ell$, with $\theta^+$ being the outer null expansion of the curve, and $L$ its length.
Since the value of $\theta^+$ depends on the normalization of the outer null normal vector $\vec{k}^+$, this has been set so that $\vec{k}^+ \cdot \vec{k}^- = -2$ and $\vec{k}^- \cdot \vec{\xi} = -1$, where $\vec{k}^-$ is the inner null normal vector to the curve and $\vec{\xi}$ is a static Killing vector field.
The type of curves to which the result applies in general are closed, smooth, non-self-intersecting, and such that past outward directed null geodesics orthogonal to the curve do never cross each other.
Standing on its own, the inequality is a curious geometric property of anti-de Sitter space.

We remark that when the curve is restricted to the hyperbolic plane, the inequality can be written as
\beq
	\frac{1}{2\pi}\oint \frac{H}{\Omega} \, \mathrm{d}l \geq 1 + \left( \frac{L}{2\pi\ell} \right)^2,
\eeq
where $H$ is the mean curvature of the curve in the hyperbolic plane, and where $\Omega$ is given by Eq. \eqref{eq:omega}.
As mentioned earlier, such curves fulfil our requirements if they are strictly geodetically convex.
This result is in perfect agreement with a special case of an inequality for (hyper)surfaces in $n$-dimensional hyperbolic space, which is proven in \cite{delima16}.

Our inequality \eqref{eq:toypi2} is saturated when the curve is a circle in the hyperbolic plane.
In this case we can give the physical interpretation that a null shell of dust has collapsed to form a nonrotating BTZ black hole, and the circle is a marginally trapped cut of the event horizon.
The connection to marginally trapped curves and black hole spacetimes makes the inequality even more interesting as an analogue of the original Penrose inequality.

In our proof, the definition of a null surface in terms of its past intersection with conformal infinity has proven beneficial.
It would be interesting to see if a similar approach can be applied in 3+1 dimensions.

\subsection*{Acknowledgements}
The authors thank Lars Andersson for inspiration and encouragement.
We thank Jos\'e M. M. Senovilla for encouragement and clarifying discussions.
Interesting and helpful comments have also been given by two anonymous reviewers.

\end{document}